\documentstyle[psfig,prl,aps]{revtex} 

\def \fij{f_{ij}}
\def \fsij{\fij^{self}}
\def \flij{\fij^{load}}
\def \eij{\epsilon_{ij}}
\def \kij{k_{ij}}

\begin{document}
\begin{twocolumn}
\title{Isostatic phase transition and instability in stiff granular materials.}
\author{Cristian F.~Moukarzel\footnote{\bf cristian@if.uff.br}}
\address{Instituto de F\'\i sica, Universidade Federal Fluminense, 
\\ 24210-340 Niter\'oi RJ, Brazil.}
\maketitle
\begin{abstract}
Structural rigidity concepts are used to understand the origin of
instabilities in granular aggregates.  It is first demonstrated that the
contact network of a noncohesive granular aggregate becomes \emph{exactly
isostatic} when $I=k\epsilon/f_l >>1$, where $k$ is stiffness, $\epsilon$ is
the typical interparticle gap and $f_L$ is the typical stress induced by
loads. Thus random packings of stiff particles are typically
isostatic. Furthermore isostaticity is responsible for the anomalously large
susceptibility to perturbation observed in granular aggregates. The
load-stress response function of granular piles is \emph{critical} (power-law
distributed) in the isostatic limit, which means that slight overloads will
produce internal rearrangements. 
\end{abstract}

Photoelastic visualization experiments~\cite{PE,Liu,Review} show clearly
defined stress-concentration paths in non-cohesive granular materials under
applied load. These often suffer sudden rearrangement on a global scale when
the load conditions are slightly changed, evidencing a degree of
susceptibility to perturbation not usually present in elastic materials. It is
rather possible that this intrinsic instability be responsible for much of the
interesting phenomenology of granular materials~\cite{Review,Book}. Recently a
number of phenomenological models~\cite{Liu,qmodel,vmodel,archmodel,Nico} have
been put forward, which succeed to reproduce several aspects of stress
propagation in granular systems, and the issue of instability has been
addressed by noting that the load-stress response function may take negative
values~\cite{vmodel}.  It is the purpose of this letter to show that
structural rigidity concepts help us understanding the \emph{origin} of
instability in granular materials, linking it to the topological properties of
the system's contact network.

Structural rigidity~\cite{Crapo} studies the conditions that a network of
points connected by rotatable bars (representing central forces) has to fulfill
in order to sustain applied loads. A network with too few bars is
\emph{flexible}, while if it has the minimum number required to be rigid it is
\emph{isostatic}. Networks with bars in excess of minimal rigidity are
\emph{overconstrained}, and are in general \emph{self-stressed}. Concepts from
structural rigidity were first introduced in the study of granular media by
Guyon et al~\cite{Guyon}, who stressed that granular systems are not entirely
equivalent to linear elastic networks since in the former only compressive
interparticle forces are possible. We next show that this constraint has
far-reaching consequences for the static behavior of stiff granular
aggregates.

Consider a $d$-dimensional frictionless granular pile in equilibrium under the
action of external forces $\vec F_i$ (gravitational, etc) on its
particles. Imagine building an equivalent linear-elastic central-force network
(the \emph{contact network}), in which two sites are connected by a bond if
and only if there is a nonzero compression force between the two corresponding
particles. Because of linearity, stresses $\fij$ on the bonds of
this equivalent system can be decomposed as $\fij=\fsij+\flij$ where $\fsij$
are self-stresses, and $\flij$ are load-dependent stresses. These last are
linear in the applied load and are do not change if all stiffnesses are
rescaled.  Self-stresses in turn do not depend on the applied load, but are
linear combinations of terms of the form $\kij \eij$ where $\kij$ are the
stiffnesses of the bonds, and $\eij$ their length-mismatches. In a granular
pile, length-mismatches are due to interparticle gaps, and therefore will
depend on the distribution of radii and on the characteristics of the
packing. Furthermore, self-stresses can only arise within
\emph{overconstrained} subgraphs~\cite{Crapo,Guyon}, i.e. those with more
contacts than strictly necessary to be rigid. It is easy to see that a bounded
overconstrained subgraph with nonzero self-stresses must have at least one
negative (traction) self-stress. It suffices to consider a joint belonging to
the envelope of the overconstrained cluster: since bonds can only reach it
from one side of the frontier, stresses of both signs are necessary in order
for the joint to be equilibrated.  Now rescale all stiffnesses according to
$k\to \lambda k$. In doing so, self-stresses are rescaled by $\lambda$,
but load-dependent stresses remain constant. Thus if self-stresses were
non-zero, in the limit $\lambda \to \infty$ at least one bond of the network
would have negative \emph{total} stress, which is a contradiction. Therefore
stiff granular piles must must either: a) have zero length-mismatches, or b)
have no overconstrained graphs at all.  Condition a) cannot be satisfied if
the particles have random polydispersity, \emph{no matter how
small}, or if the packing is disordered. 
Therefore there can be no overconstrained subgraphs in polydisperse or
disordered packings in the large-stiffness limit. In other words: \emph{ the
contact network of a granular packing becomes isostatic when the stiffness is
so large that the typical self-stress, which is of order $k\epsilon$, would be
much larger than the typical load-induced stress $f_L$~\cite{exceptions}}.

The isostaticity condition above is perhaps simpler to understand when cast in 
the following terms: granular packings will only fail to be isostatic if the
applied compressive forces are strong enough to close interparticle gaps
establishing redundant contacts.

Thus, real disordered packings will typically be isostatic (interparticle gaps
are large) unless its particles are strongly deformed by the load. This
explains why the average coordination number of random sphere packings is
usually close to 6 (see \cite{Guyon} and references therein). Isostaticity was
also reported in numerical simulations of rigid disks~\cite{OR}.

Finally we note that an adimensional ``isostaticity parameter'' can be defined
as $I= k \epsilon / f_L$, and that $I>>1$ corresponds to the isostatic limit.
\vbox{
\begin{figure}[tbhp]
\centerline{\psfig{figure=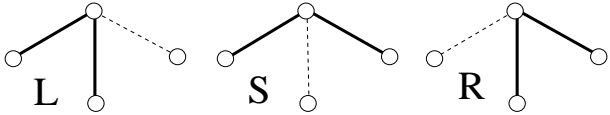,width=8cm,angle=270}}
\centerline{}
\caption{  
Appropriately choosing among these three isostatic configurations 
for each site, only compressive stresses are produced on a triangular
packing. First S is chosen with probability 1/2. If S is not chosen 
then either R or L are, depending on the sign of the horizontal 
force acting on the site.}
\label{fig:1} 
\end{figure} }
We now discuss the consequences of isostaticity for the static behavior of a
pile. It is possible to obtain useful insight from recent studies of the
related problem of central-force rigidity percolation~\cite{Lett1}.  Rigid
backbones are found to be composed of large overconstrained clusters,
\emph{isostatically} connected to each other by \emph{critical bonds} (also
called red bonds). Cutting one critical bond is enough to produce the collapse
of the entire system, because each of these is by definition essential for
rigidity. In percolation backbones though, the number of such critical bonds
is not extensive, but scales at $p_c$ as $L^{1/\nu}$ where $\nu$ is the
correlation-length exponent~\cite{Coniglio}.  Thus, if we perturb (cut or
stretch) a randomly chosen bond in a percolation backbone, most of the times
the effect will be only be local since no critical bond will be hit. The new
element in stiff granular contact-networks is the fact that \emph{all}
contacts are isostatic, or critic, i.e. there is \emph{extended
isostaticity}. Thus we may expect stiff granular systems to have a large
susceptibility to perturbation since cutting (stretching) a bond will often
produce a large part of the system to collapse (move).

Let us now quantify these ideas. We perturb the system by introducing an
\emph{infinitesimal} change $\delta l$ in the length of a randomly chosen
bond, and record the \emph{induced displacement} $\delta \vec x_i$ suffered by
particle centers in equilibrium. The system's susceptibility to perturbation
is then defined as \hbox{$D=\sum_{i=1}^N { | \delta \vec x_i / \delta l|}^2$}.
We propose to measure $D(O_v)$ as a function of the density $O_v$ of
\emph{overconstraints} (excess contacts) randomly located on the network.
Isostatic piles have $O_v=0$.

A simple one-dimensional model for the propagation of perturbations can be
analytically solved\cite{tbp} for arbitrary values of the density $O_v$ of
overconstraints. For any non-zero $O_v$, $D$ as defined above takes a finite
value, but diverges as $O_v^{-1/2}$ for $O_v \to 0$. Therefore there is a
phase transition at the isostatic point $O_v=0$.

We now analyze a two dimensional system numerically. In the spirit of
previously studied models~\cite{Roux}, we consider a triangular packing of $H$
layers height, with one of its principal axis parallel to gravity, and made of
disks with small random polydispersity $\delta R$ and weight $W$.  Since
$\delta R$ is small, disk centers are approximately located on the sites of a
regular triangular lattice. If the stiffness $k$ is large enough ($I=k \delta
R / W H >>1$), the contact network will be isostatic. We enforce isostaticity
in our model by letting each site be supported from below by only two out of
its three neighbors.  This gives three possible local configurations which are
depicted in Fig.\ref{fig:1}. By appropriately choosing among these~\cite{tbp},
random isostatic networks with only compressive stresses are generated. In
order to study the effect of a finite density $O_v$ of overconstraints (which
would appear if the stiffness is lowered), we furthermore let \emph{all three
bonds} be present with probability $O_v$ at each site. 

After building a disordered network in this way, a randomly chosen bond in the
lowest layer is stretched, and the induced displacement field is
measured. After averaging over disorder, the stress distribution~\cite{tbp} is
found to decay exponentially for large stresses, in accordance with previous
work~\cite{Liu,Radjai}. 

The results for the susceptibility $D_y(H,O_v) = \sum_{i=1}^N {\delta y_i}^2$
are shown in Fig.~\ref{fig:2}a. Here $\delta y_i$ is the vertical displacement
of site $i$ due to a unitary bond-stretching, as measured on packings of $N=H \times
H$ particles.  For $O_v > 0$, $D_y$ goes to a \emph{finite limit} for large
sizes $H$, but diverges with system size if $O_v=0$. Measurements on isostatic
packings of up to $H=2000$ layers~\cite{tbp} show that $\log{D_y} \propto H$,
i.e. the divergence of $D_y$ is exponential with size when $O_v=0$. Thus there
is a surprising \emph{phase transition} at $O_v=0$, where anomalously large
susceptibility sets in. 

In order to understand how displacements propagate upwards, we measure the
probability distribution $P_h(\delta y)$ to have a vertical displacement
$\delta y$, $h$ layers above the perturbation. Numerical results for isostatic
systems with $H=2000$ are shown in Fig.~\ref{fig:2}b. For large $h$,
$P_h(\delta y)$ decays as a power-law with an $h$-dependent cutoff:
$P_h(\delta y) \sim h^{-\rho}|\delta y|^{-\theta}$, $\delta y< \delta_M(h)$.
As seen in Fig.~\ref{fig:2}b, $\delta_M(h)$ grows exponentially with height
$h$. This produces the observed exponential divergence of
$D_y$. Similar measurements were done on systems with a finite density of
overconstraints $O_v$, in which case the distribution of displacements
presents a height-independent bound~\cite{tbp}.

The puzzling appearance of exponentially large displacements on isostatic
piles can be explained as due to the existence of ``lever configurations'' or
``pantographs'', which amplify displacements.  Fig.~\ref{fig:3} shows an
example of a pantograph with amplification factor 2. Given that this and
similar mechanisms appear with a finite density per layer, it is clear that
the second moment of $P_h(\delta y)$ will grow exponentially with system
height.  Furthermore, it is easy to understand why this amplification effect
only exists in the \emph{isostatic limit}: Pantographs as the one in
Fig. \ref{fig:3} are no longer effective if blocked by overconstraints, for
example if an additional bond is added between site A and the site below
it. In this case, a stretching of bond B would induce stresses in the whole
pantograph, and only a small displacement of site A.
\vbox{
\begin{figure}[tbhp]
\vbox{
\centerline{\psfig{figure=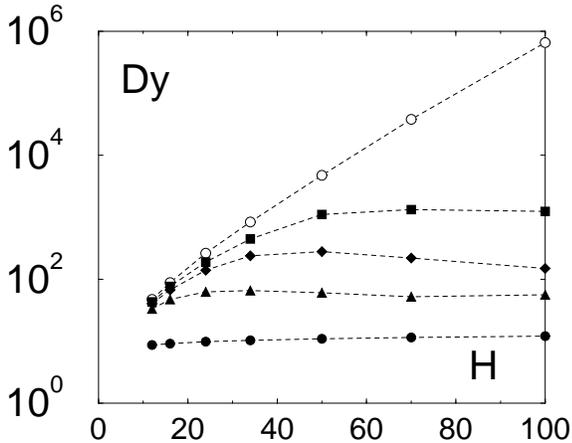,width=8cm,angle=270}}
\centerline{\psfig{figure=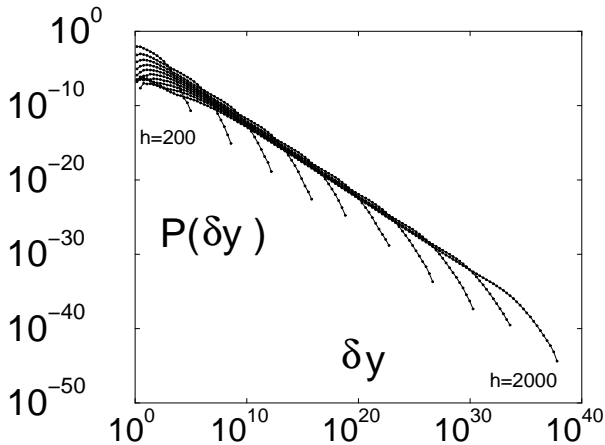,width=8cm,angle=270}}
}
\centerline{}
\caption{ 
{ \bf a)}Total susceptibility $D_y$ versus system height $H$ in
layers, as numerically measured on two-dimensional triangular packings. 
The fraction of overconstraints (fraction of sites supported by three
lower neighbors) is: $0.00$ (open circles), $0.01$ (squares), $0.02$
(diamonds), $0.05$ (triangles) and $1.00$ (full circles).
{\bf b)} The probability
$P_H(\delta y)$ to have an induced vertical displacement $\delta y$, $H$
layers above the perturbation, as obtained in a numerically exact fashion
for the isostatic ($O_v=0$) triangular piles described in the text.
Results are shown for $H=200,400,600,\ldots,2000$. Only positive values
of $\delta y$ have been plotted here.
}
\label{fig:2}
\end{figure} }
In order to formalize the relationship between these findings and the observed
unstable behavior~\cite{PE,Liu} of granular materials, we now demonstrate the
equivalence between induced displacements and the
load-stress response function~\cite{vmodel} of the stretched bond.  The
network's total energy can be written as \hbox{$E=\sum_{i=1}^N W_i y_i + 1/2
\sum_{b} k_b (l_b - l^0_b)^2$} where the first term is the potential energy
and the second one is a sum over all bonds and accounts for the elastic
energy.  $l_b$ are bond lengths in equilibrium and $l^0_b$ their repose
lengths. Upon infinitesimally stretching bond $b'$, equilibrium requires that
$
\sum_i W_i \frac{\partial y_i}{\partial l_{b'}} + \sum_{ov} k_{ov} 
( l_{ov} - l_{ov}^0) \frac{\partial
l_{ov}}{\partial l_{b'}} = 0
$, 
where the second sum goes over bonds $ov$ that belong to the same
\emph{overconstrained} graph as $b'$ does. This is so since bonds not
overconstrained with respect to $b'$ \emph{do not change their lengths} as a
result of stretching $b'$. Since stress $f_b$ on bond $b$ is
$f_b=k_b(l_b^0-l_b)$ this may be rewritten as
\begin{equation}
 \sum_{ov} f_{ov} \frac{\partial l_{ov}}{\partial l_{b'}} =
\sum_i W_i \frac{\partial y_i}{\partial l_{b'}} 
\label{eq:response}
\end{equation}
If $b'$ does not belong to an overconstrained graph, the left hand sum only
contains bond $b'$ itself, therefore \hbox{$f_{b}= \sum_i W_i \frac{\partial
y_i}{\partial l_{b}}$} showing that, in the isostatic case (no overconstrained
graphs at all), the induced displacement \hbox{$\delta y_i^{(b)}=
\frac{\partial y_i}{\partial l_b}$} is equal to the \emph{response function}
of stress $f_b$ with respect to an overload on site $i$.
\vbox{
\begin{figure}[tbhp]
\centerline{ \psfig{figure=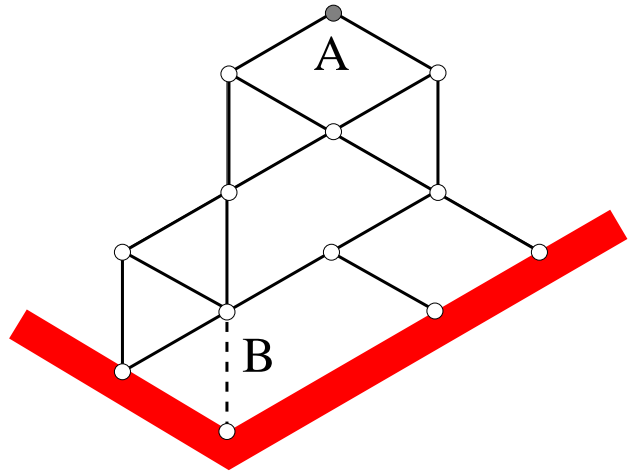,width=8cm,angle=270}}
\centerline{}
\caption{ 
The observed exponential growth of induced displacements is due to
the existence of random ``pantographs'' as the one shown in this figure. Upon
stretching bond B by an amount $\delta$, site A moves vertically
by an amount $2\delta$. Conversely a unitary weight at A produces
a stress of magnitude 2 on bond B. This is a consequence of the general
equivalence between induced displacements and the load-stress response
function. 
}
\label{fig:3}
\end{figure} }
Taking averages with respect to disorder we obtain $<f_b>= \sum_i W_i <\delta
y_i^{(b)}>$ and since average stresses grow as $H$ we must have $<\delta
y_i^{(b)}>\sim H^{-1}$. We thus see that there must be delicate cancellations
in $P(\delta y)$, since its second moment diverges as $\exp\{H\}$ while its
first moment goes to zero with $H$. This shows that $\delta y$ (and therefore
the response function) takes exponentially large values of \emph{both signs},
($P(\delta y)$ is approximately symmetric). Thus a positive overload at site
$i$ would often produce a (very large) negative stress on bond $b$, implying
the need for rearrangement since negative stresses are not allowed.

The existence of negative values for the response function was first discussed
in relation with instability, in the context of a phenomenological vectorial 
model for stress propagation~\cite{vmodel}. The results of the present work
demonstrate that the response function takes exponentially large negative
values, and the system is unstable, because of the isostatic character of the
contact network.  

To conclude, we have shown that granular packings are \emph{exactly isostatic}
when $I = k \epsilon/ f_L$ is much larger than one,
which holds for typical disordered packings, and also for stiff enough regular
packings with random polydispersity. 

For isostatic packings, the distribution of displacements induced by a
perturbation is power-law with an exponentially large cutoff.  A
susceptibility to perturbation can be defined, which diverges upon increasing
$I$. Thus, an isostatic phase transition takes place in the limit of large
$I$.

Induced displacements were furthermore shown to be equivalent to the
load-stress response function of the perturbed bond.  Our results for induced
displacements thus mean that response functions take exponentially large
values, as well positive as negative, in the isostatic limit. This explains
why stiff granular piles are unstable. Any non-zero density of overconstraints
destroys criticality and therefore instabilities will not be present when the
isostaticity parameter $I$ is small. $I$ can be reduced by reducing the
stiffness, the interparticle gaps, or by increasing the load.
\acknowledgements 
I thank H.~J.~Herrmann, J.~L.~Goddard P.~M.~Duxbury and J.~Socolar for useful
discussions and comments, and H.~Rieger for an important suggestion. HLRZ
J\"ulich provided supercomputer access and warm hospitality.
The author is supported by CNPq, Brazil.
%
%

\end{twocolumn}


\begin{references}
\bibitem{exceptions}

	Exceptions are packings with periodic boundary conditions because
	they are not bounded, and idealized regular close-packings, because
	they have no lenght-mismatches and therefore no self-stress even on
	overconstrained graphs.
\end{references}
\end{document}